\documentclass[aps,prd,twocolumn,superscriptaddress,showpacs,amsmath,amssymb,floatfix]{revtex4-1}

\usepackage[pdfauthor={Marko Petric}]{hyperref}
\usepackage{graphicx}
\usepackage{dcolumn}
\usepackage{bm}
\usepackage{url}
\usepackage{relsize}

\RequirePackage{xspace}

\newcommand{\gev}{\ensuremath{{\mathrm{\,Ge\kern -0.1em V\!}}}\xspace}
\newcommand{\mev}{\ensuremath{{\mathrm{\,Me\kern -0.1em V\!}}}\xspace}
\newcommand{\gevc}{\ensuremath{{\mathrm{\,Ge\kern -0.1em V\!/}c}}\xspace}
\newcommand{\mevc}{\ensuremath{{\mathrm{\,Me\kern -0.1em V\!/}c}}\xspace}
\newcommand{\gevcc}{\ensuremath{{\mathrm{\,Ge\kern -0.1em V\!/}c^2}}\xspace}
\newcommand{\mevcc}{\ensuremath{{\mathrm{\,Me\kern -0.1em V\!/}c^2}}\xspace}

\newcommand{\dz}{\ensuremath{D{}^0}\xspace}
\newcommand{\ds}{\ensuremath{D{}_{s}}\xspace}
\newcommand{\dsp}{\ensuremath{D{}_{s}^{+}}\xspace}
\newcommand{\dpms}{\ensuremath{D^{*+}}\xspace}
\newcommand{\brll}[2]{\mathcal{B}\left(\dz\to#1^{+}#2^{-}\right)}

\begin{document}  

\title{Search for leptonic decays of $D^0$ mesons}

\affiliation{Budker Institute of Nuclear Physics, Novosibirsk}
\affiliation{Faculty of Mathematics and Physics, Charles University, Prague}
\affiliation{University of Cincinnati, Cincinnati, Ohio 45221}
\affiliation{Justus-Liebig-Universit\"at Gie\ss{}en, Gie\ss{}en}
\affiliation{Hanyang University, Seoul}
\affiliation{University of Hawaii, Honolulu, Hawaii 96822}
\affiliation{High Energy Accelerator Research Organization (KEK), Tsukuba}
\affiliation{Institute of High Energy Physics, Chinese Academy of Sciences, Beijing}
\affiliation{Institute of High Energy Physics, Vienna}
\affiliation{Institute of High Energy Physics, Protvino}
\affiliation{Institute for Theoretical and Experimental Physics, Moscow}
\affiliation{J. Stefan Institute, Ljubljana}
\affiliation{Kanagawa University, Yokohama}
\affiliation{Institut f\"ur Experimentelle Kernphysik, Karlsruhe Institut f\"ur Technologie, Karlsruhe}
\affiliation{Korea University, Seoul}
\affiliation{Kyungpook National University, Taegu}
\affiliation{\'Ecole Polytechnique F\'ed\'erale de Lausanne (EPFL), Lausanne}
\affiliation{Faculty of Mathematics and Physics, University of Ljubljana, Ljubljana}
\affiliation{University of Maribor, Maribor}
\affiliation{Max-Planck-Institut f\"ur Physik, M\"unchen}
\affiliation{University of Melbourne, School of Physics, Victoria 3010}
\affiliation{Nagoya University, Nagoya}
\affiliation{Nara Women's University, Nara}
\affiliation{National Central University, Chung-li}
\affiliation{National United University, Miao Li}
\affiliation{Department of Physics, National Taiwan University, Taipei}
\affiliation{H. Niewodniczanski Institute of Nuclear Physics, Krakow}
\affiliation{Nippon Dental University, Niigata}
\affiliation{Niigata University, Niigata}
\affiliation{Novosibirsk State University, Novosibirsk}
\affiliation{Osaka City University, Osaka}
\affiliation{Panjab University, Chandigarh}
\affiliation{University of Science and Technology of China, Hefei}
\affiliation{Seoul National University, Seoul}
\affiliation{Sungkyunkwan University, Suwon}
\affiliation{School of Physics, University of Sydney, NSW 2006}
\affiliation{Tata Institute of Fundamental Research, Mumbai}
\affiliation{Excellence Cluster Universe, Technische Universit\"at M\"unchen, Garching}
\affiliation{Tohoku Gakuin University, Tagajo}
\affiliation{Tohoku University, Sendai}
\affiliation{Department of Physics, University of Tokyo, Tokyo}
\affiliation{Tokyo Metropolitan University, Tokyo}
\affiliation{Tokyo University of Agriculture and Technology, Tokyo}
\affiliation{IPNAS, Virginia Polytechnic Institute and State University, Blacksburg, Virginia 24061}
\affiliation{Yonsei University, Seoul}
  \author{M.~Petri\v c}\affiliation{J. Stefan Institute, Ljubljana} 
  \author{M.~Stari\v c}\affiliation{J. Stefan Institute, Ljubljana} 
  \author{I.~Adachi}\affiliation{High Energy Accelerator Research Organization (KEK), Tsukuba} 
  \author{H.~Aihara}\affiliation{Department of Physics, University of Tokyo, Tokyo} 
  \author{K.~Arinstein}\affiliation{Budker Institute of Nuclear Physics, Novosibirsk}\affiliation{Novosibirsk State University, Novosibirsk} 
  \author{T.~Aushev}\affiliation{\'Ecole Polytechnique F\'ed\'erale de Lausanne (EPFL), Lausanne}\affiliation{Institute for Theoretical and Experimental Physics, Moscow} 
  \author{A.~M.~Bakich}\affiliation{School of Physics, University of Sydney, NSW 2006} 
  \author{V.~Balagura}\affiliation{Institute for Theoretical and Experimental Physics, Moscow} 
  \author{E.~Barberio}\affiliation{University of Melbourne, School of Physics, Victoria 3010} 
  \author{K.~Belous}\affiliation{Institute of High Energy Physics, Protvino} 
% ...
\collaboration{The Belle Collaboration}
\noaffiliation

\date{\today}

\begin{abstract} 
  We search for the flavor-changing neutral current decays 
  $D^0 \rightarrow \mu^+\mu^-$ and $D^0 \rightarrow e^+e^-$, and for the lepton-flavor violating
  decays $D^0 \rightarrow e^\pm\mu^\mp$ using 660 fb${}^{-1}$ of data collected with the Belle detector 
  at the KEKB asymmetric-energy $e^{+}e^{-}$ collider. We find no evidence for any of these decays. We obtain significantly improved upper limits on the branching fractions: $\mathcal{B}\left(D^0 \rightarrow \mu^{+}\mu^{-}\right)<1.4\times10^{-7},~\mathcal{B}\left(D^0 \rightarrow e^{+}e^{-}\right)<7.9\times 10^{-8}$ and $\mathcal{B}\left(D^0 \rightarrow e^{+}\mu^{-}\right)+\mathcal{B}\left(D^0 \rightarrow \mu^{+}e^{-}\right)<2.6\times 10^{-7}$ at 90\% confidence level.
\end{abstract}

\maketitle    

\begin{figure}[htp]
\includegraphics[width=0.45\columnwidth]{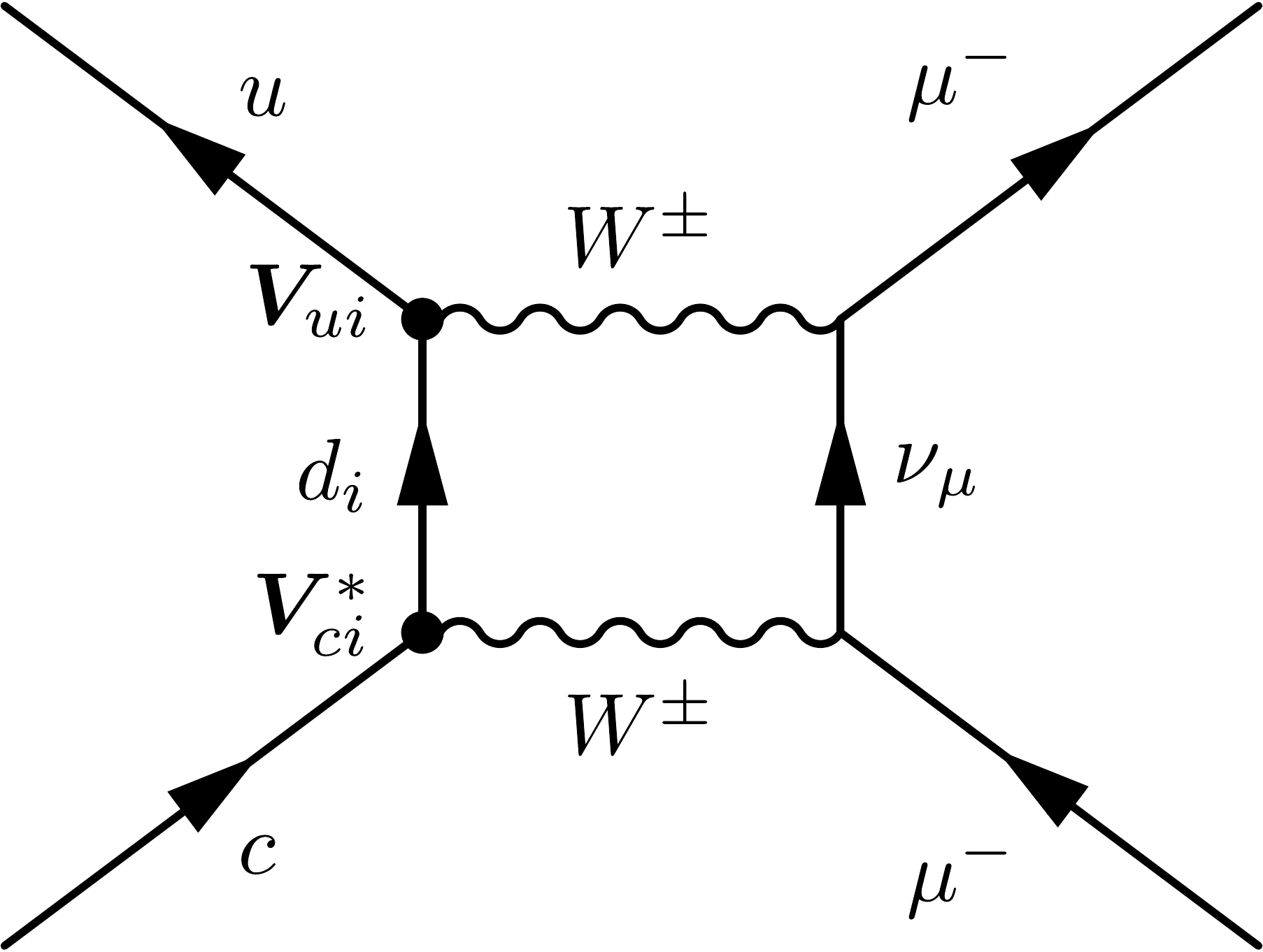}\hspace{0.2cm}%
\includegraphics[width=0.45\columnwidth]{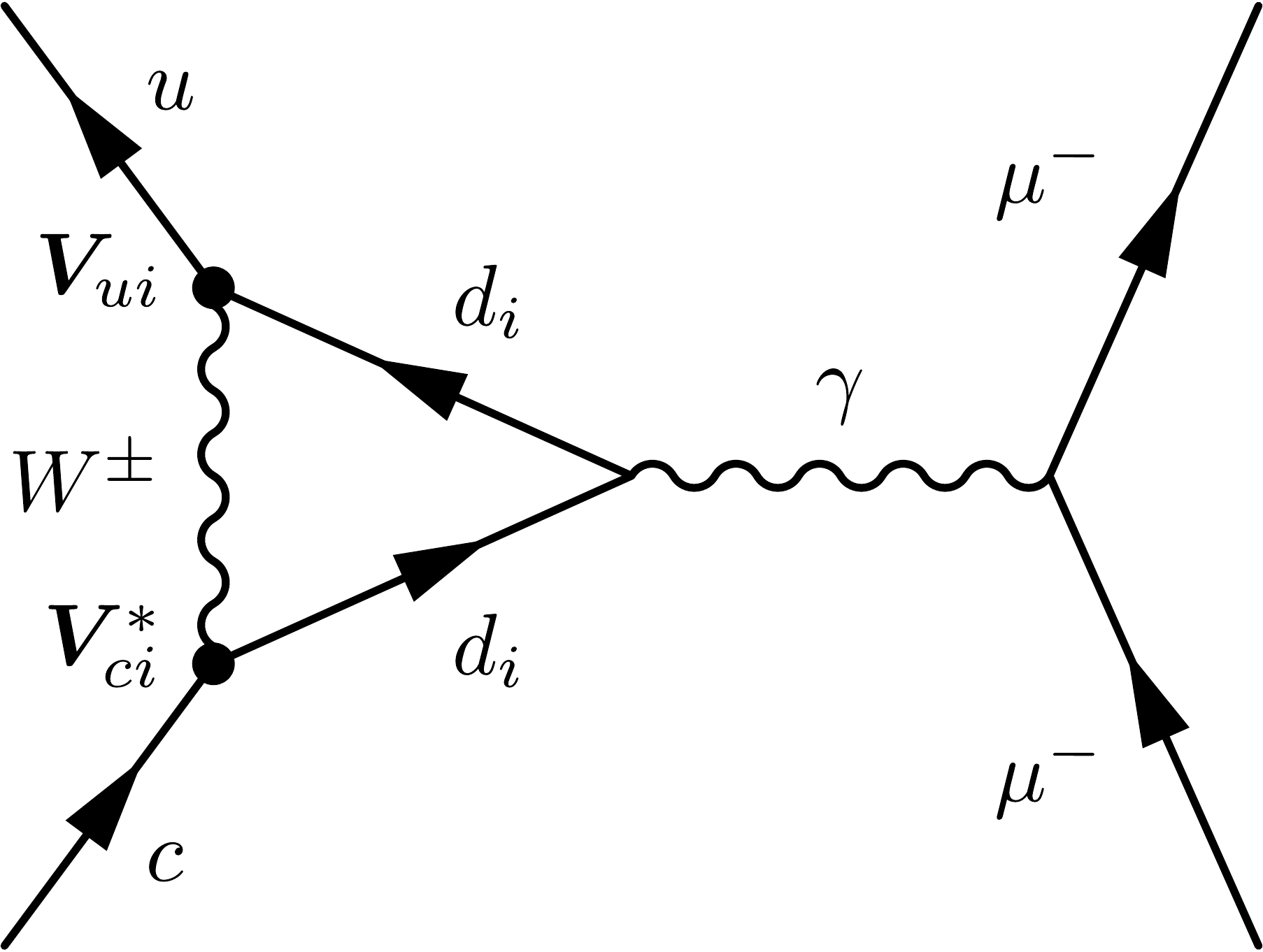}
\caption{The SM short distance Feynman diagrams for the $\dz \to \mu^{+}\mu^{-}$ decay.
}
\label{fig:Dtomumu} 
 \end{figure}

The flavor-changing neutral current (FCNC) decays $\dz\to~e^+e^-$ and 
$\dz\to\mu^+\mu^-$~\cite{cc} are highly suppressed in the standard model (SM) by the 
Glashow-Iliopoulos-Maiani mechanism~\cite{gim}. With the inclusion of long distance contributions the branching fractions can reach values of around $10^{-13}$~\cite{smfcnc}. The SM short distance Feynman diagrams for the $\dz \to \mu^{+}\mu^{-}$ decay are shown in Fig.~\ref{fig:Dtomumu}. The lepton-flavor violating (LFV) decays $\dz\to e^\pm\mu^\mp$ are forbidden in the SM, but are possible in extensions of the SM with nondegenerate neutrinos and nonzero neutrino mixings and are expected to be of the order of $10^{-14}$~\cite{smfcnc} in these scenarios. All these predictions are orders of magnitude 
below the current experimental sensitivity.

In certain new physics (NP) scenarios, FCNC branching fractions can be enhanced by 
many orders of magnitude. For example, $R$-parity violating supersymmetry can 
increase the branching fractions of $\dz\to~e^+e^-$ and $\dz\to~\mu^+\mu^-$ 
up to $10^{-12}$ and $10^{-8}$, respectively~\cite{newphys}.
The latter prediction 
is close to the current experimental sensitivity. 
As another example, so far unobserved leptoquarks were suggested as a possible explanation of the small discrepancy between the measured value of the $\ds$ meson decay constant and the prediction of lattice QCD~\cite{dobrescu:241802}. Leptoquarks could also enhance the $\dz \to \ell^{+}\ell^{-}$ branching fraction. In order to explain the measured $\dsp\to\mu^{+}\nu$ width by a leptoquark contribution, and comply with other constraints arising from charm meson decays, $\brll{\mu}{\mu}$ should be enhanced to $8\times 10^{-7}$~\cite{Dorsner:2009cu}. 
The above examples demonstrate the importance of FCNC and LFV decays searches in the exploration of possible NP contributions. It should be noted that charm FCNC and LFV decays probe the couplings of the up-quark sector in contrast to $B$ or $K$ meson decays. 

In this paper, we report on a search for the decays $\dz\to\mu^+\mu^-$, 
$\dz\to e^+e^-$ and $\dz\to e^\pm\mu^\mp$ using $660\,\mbox{fb}^{-1}$ of data recorded in $e^+e^-$ collisions at the center-of-mass (CM) energy of the $\Upsilon(4S)$ resonance and 60~\mev 
below by the Belle detector at the KEKB collider.

The Belle detector, which is described in detail elsewhere~\cite{Belle}, 
is a large-solid-angle magnetic spectrometer that consists of a silicon 
vertex detector (SVD)~\cite{svd}, a 50-layer central drift chamber (CDC), 
an array of aerogel threshold Cherenkov counters (ACC), a barrel-like 
arrangement of time-of-flight scintillation counters (TOF), and an 
electromagnetic calorimeter composed of CsI(Tl) crystals (ECL) located 
inside a superconducting solenoid coil that provides a $1.5\,$T magnetic 
field. An iron flux-return located outside the coil is instrumented to 
detect $K_L^0$ mesons and to identify muons (KLM). Two inner detector 
configurations were used. A beam pipe with a radius of  $2.0\,$cm and a 3-layer silicon 
vertex detector were used for the first sample of
$155\,\mbox{fb}^{-1}$, while a $1.5\,$cm beampipe, a 4-layer silicon 
detector, and a small-cell inner drift chamber were used to record the 
remaining data sample.

In this measurement only $\dz$ mesons coming from $c$ - quark production in 
the continuum $e^{+}e^{-}\to c \overline{c}$ process are considered. The inclusion of 
$\dz$ mesons from $B$ decays would result in a higher combinatorial background. We normalize the sensitivity of our search to topologically similar $\dz\to\pi^{+}\pi^{-}$ decays; this cancels various systematic uncertainties.
The $\dz\to\ell^+\ell^-\;(\ell=e~\mathrm{or}~\mu)$ branching fraction is determined by 
\begin{equation} 
  \brll{\ell}{\ell} = 
  N_{\ell\ell} f\label{breq} 
\end{equation} 
where $N_{\ell\ell}$ is the number of reconstructed $\dz\to\ell^+\ell^-$ decays
and $f$ is defined as 
\begin{equation} 
  f \equiv 
  \frac{1}{N_{\pi\pi}} 
  \frac{\epsilon_{\pi\pi}}{\epsilon_{\ell\ell}} 
  \brll{\pi}{\pi}
  \label{feq} 
\end{equation} 
Here $\brll{\pi}{\pi}=(1.397\pm0.027)\times 10^{-3}$ is the well-measured
$\dz\to\pi^+\pi^-$ branching fraction~\cite{PDG}, $N_{\pi\pi}$ is the 
number of reconstructed $\dz\to\pi^+\pi^-$ decays, and $\epsilon_{\ell\ell}$ and
$\epsilon_{\pi\pi}$ are the reconstruction efficiencies for
$\dz\to\ell^{+}\ell^{-}$ and $\dz\to \pi^{+}\pi^{-}$ decays, respectively.

First, a general event selection is performed that is, apart from the particle
identification criteria, the same for all data 
samples. Later in the analysis, tighter optimized criteria specific for each decay mode are used.

We use $\dz$ mesons from the decay $\dpms\to\dz\pi^{+}_{s}$ with a characteristic low momentum pion, since this considerably improves the purity of the $\dz\to \ell^+\ell^-$ and $\pi^+\pi^-$ samples. Each charged track is required to have at least two associated vertex detector hits in each of the two measurement coordinates. To select pion and lepton candidates, we 
impose standard particle identification criteria. Charged pions are identified using $\mathrm{d}E/\mathrm{d}x$ measurement from the 
CDC, Cherenkov light yields in the ACC, and timing information from the TOF~\cite{PID}. Muon identification is based on the matching quality and penetration depth of associated
hits in the KLM~\cite{muid}. Electron identification is determined using the ratio
of the energy deposit in the ECL to the momentum measured in the SVD
and CDC, the shower shape in the ECL, the matching between the
position of charged track trajectory and the cluster position in the
ECL, the hit information from the ACC and the $\mathrm{d}E/\mathrm{d}x$ information in
the CDC~\cite{elid}. The muon and 
electron identification efficiencies are around 90\% with less than 1.5\% and 0.3\%
pion misidentification, respectively, whereas the pion 
identification efficiency is around 83\%. $\dz$ daughter tracks are 
refitted to a common vertex, and the $\dz$ production vertex is found by 
constraining the $\dz$ trajectory and the $\pi_s$ track to originate from 
the $e^+e^-$ interaction region; confidence levels exceeding $10^{-3}$ are 
required for both fits. A $\dpms$ momentum greater than $2.5\,\gev/c$ in 
the CM frame of the collisions is required to reject $D$-mesons produced in 
$B$-meson decays and to suppress combinatorial background.

Candidate $\dz$ mesons are selected using two kinematic observables: the 
invariant mass of the $\dz$ decay products, $M$, and the energy released 
in the $\dpms$ decay, $q=(M_{\dpms}-M-m_\pi)c^2$, where $M_{\dpms}$ is the 
invariant mass of the $\dz \pi_s$ combination and $m_\pi$ is the $\pi^+$ 
mass~\cite{PDG}. We require $1.81\gevcc<M<1.91\gevcc$ and $q<20\mev$.

According to Monte Carlo (MC) simulation based on {\sc EvtGen}~\cite{evtgen} and
{\sc GEANT3}~\cite{geant}, the background in $\dz \to \ell^{+}\ell^{-}$ decays originates 
predominantly from semileptonic $B$ decays (80\%) and from $\dz$ 
decays (10\%). 
The background events can be grouped into two categories based on their $M$ distribution: 
(1) a smooth combinatorial background, and (2) a 
peaking background from the misidentification of $\dz\to \pi^+\pi^-$ 
decays. The decay $\dz\to \pi^+\pi^-$ contributes to the background 
when both pions are misidentified as muons or as a muon-electron combination. MC studies showed 
that misidentification of a $K$ meson as a lepton does not produce a peak inside the mass region 
$1.81\gevcc<M<1.91\gevcc$. 

The signal efficiencies $\epsilon_{\ell\ell}$ and $\epsilon_{\pi\pi}$ are evaluated using signal MC simulation, which is also based on {\sc EvtGen} and {\sc GEANT3}, but includes final-state radiative effects (FSR) simulated by {\sc PHOTOS}~\cite{photos}. Since we find differences between the widths of the $\dz\to\pi^{+}\pi^{-}$ signal in the data and MC simulation, we perform fits to the $M$ and $q$ distributions to obtain scaling factors for the signal widths, and then tune the shapes in the MC simulation by correcting $M$ and $q$ for each MC signal event by $M' = m_{0} + (M-m_{0})f_{m}$ and $q' = q_{0}+(q - q_{0})f_{q}$. Here, $m_{0}$ and $q_{0}$ denote the nominal $\dz$ mass and the nominal energy released in the $\dpms$ decay, respectively, and $f_{m} = 1.17$ and $f_{q} = 1.28$ are the corresponding scaling factors. Another difference, a small shift of $-0.2\gev$ in the mean of the distribution of the missing energy of the
event ($E_{\rm miss}$) between data and MC simulation is observed; we correct the MC distribution by subtracting, for every signal event, the above value from its $E_{\rm miss}$. We construct $E_{\mathrm{miss}}$ from the difference between the beam energy and the sum of the energies of all four vectors of photons and charged tracks, which are assumed to be pions. 
The constants derived from $\dz \to \pi^{+}\pi^{-}$ are used to correct $\dz \to \ell^{+}\ell^{-}$ MC events. The uncertainties of the tuning procedure are included in the systematic error. 

In order to avoid biases, a blind analysis technique has 
been adopted. All events inside the $\dz$ signal region of $|\Delta M|<20 \mevcc$ and $|\Delta q|<1\mev$ were blinded
 until the final event selection criteria were established. Since $\dz \to \ell^{+}\ell^{-}$ decays are not expected to be observed at the current sensitivity, we optimize the selection criteria to obtain the best upper limits;  we maximize the figure of merit, 
$\mathcal{F}=\epsilon_{\ell\ell}/N_{\rm UL}$, where $\epsilon_{\ell\ell}$ is the efficiency for 
detecting $D^0\to \ell^+\ell^-$ decays obtained from the tuned signal MC simulation and 
$N_{\rm UL}$ is the Poisson average of 
Feldman-Cousins 90\% confidence level upper limits on the number of 
observed signal events that would be obtained with the expected background 
and no signal~\cite{fc}. The average upper limits $N_{\rm UL}$ are calculated from the number of generic MC background 
events, surviving the selection criteria and scaled to the data size. The sample corresponds to 6-times the
statistics of the data.

For the optimization we select the following variables: signal
region  size ($\Delta M, \Delta q$), $E_{\rm miss}$, and minimal lepton identification
probabilities. The quantities $\Delta M$ and $\Delta q$  are measured relative to the nominal $\dz$ mass and nominal energy released in
the $\dpms$ decay, respectively. The signal region in $M$ is allowed
to be asymmetric with respect to the nominal $\dz$ mass; for the $\mu\mu$ decay mode this provides some suppression of misidentified $\dz \to \pi^+\pi^-$ decays,
since their invariant mass distribution peaks about 2 standard deviations below the $\dz$ mass; for the $ee$ and $e\mu$ modes an asymmetric requirement accounts for the low mass 
tail due to electron bremsstrahlung. 
The requirement on the maximal allowed missing energy in the event 
is chosen to suppress background from semileptonic $B$ decays; 
these events have larger missing energy due to undetected neutrinos.
We found a broad maximum in $\mathcal{F}$
for the lepton identification probability, hence we repeated the
procedure at fixed lepton identification criteria, optimizing only the size of the signal region and the maximal allowed missing energy in an
event. The results are summarized in Table~\ref{optima}. 

\begin{table}
  \caption{\label{optima} Optimal selection criteria. The requirements on $M$ 
  are asymmetric and are given as lower and upper bounds on $\Delta M$. }
  \begin{ruledtabular}
    \begin{tabular}{cccc}
      Mode & $\Delta M $ & $\Delta q$ & $E_{\rm miss}$ \\ 
      & [\mevcc] & [\mev] & [\gev] \\
      \hline 
      $\mu^+\mu^-$   & $(-8, 19)$ & $\pm$0.48 & 1.4 \\
      $e^+e^-$       & $(-27, 14)$ & $\pm$0.40 & 1.0 \\
      $e^\pm\mu^\mp$ & $(-13, 15)$ & $\pm$0.46 & 1.0 \\
\end{tabular}
\end{ruledtabular}
\end{table}

To estimate the number of combinatorial background events in the signal region,
the sideband region $|\Delta q|>1\mev$ is used.  This region is chosen 
to reduce the statistical error and to exclude possible signal events 
and misidentification from $\dz \to \pi^+\pi^-$ decays. The comparison of data and MC simulation shows
good agreement in the combinatorial background distribution in this
region. The distribution is parametrized as 
$f(M,q)=A(1-BM)\sqrt{q}$, where the parameters $A$ and $B$ are determined from a fit
to the generic MC sample. The number of combinatorial background events in 
the signal region is calculated as $N_{\rm bkg}^{\rm comb}=p\times N_{\rm side}$,
where $N_{\rm side}$ is the number of events found in the sideband region
and $p$ is the expected ratio of events in the signal and sideband region determined 
by integration of $f(M,q)$.

The peaking background in the signal region due to misidentification 
of $\dz \to \pi^+\pi^-$ decays is estimated from the reconstructed $\dz \to \pi^+\pi^-$
decays found in data by replacing the pion mass with the lepton mass and by weighting each event by
\begin{equation}
  w=\frac{u(p_1,\cos \theta_1) u(p_2,\cos \theta_2)}
  {v(p_1,\cos \theta_1) v(p_2,\cos \theta_2)}
\end{equation}
where $p_{1,2}$ and $\theta_{1,2}$ are the momenta and polar angles of the outgoing pions and where $u$ and $v$ are the pion-lepton misidentification probability and pion
identification efficiency, respectively.  The misidentification probabilities and efficiencies are measured in data using $\dpms\to~\dz~\pi_{s}^+,\, \dz\to~K^{-}\pi^{+}$ decays, binned in particle momentum $p$ and cosine of polar angle. 

The estimates for the number of background events in  the signal
region are summarized in Table~\ref{tbl:br}. The misidentification
of $\dz \to \pi^+\pi^-$ contributes significantly only to the 
$\dz \to \mu^+\mu^-$ decay channel ($1.8$ events).
The uncertainties in the background estimates listed in Table~\ref{tbl:br}
include the statistical error on the number of sideband region events
and the uncertainties in the combinatorial background parametrization,
while the uncertainty in the peaking background estimation is
negligible.

\begin{figure}[htbp]
\includegraphics[width=\columnwidth]{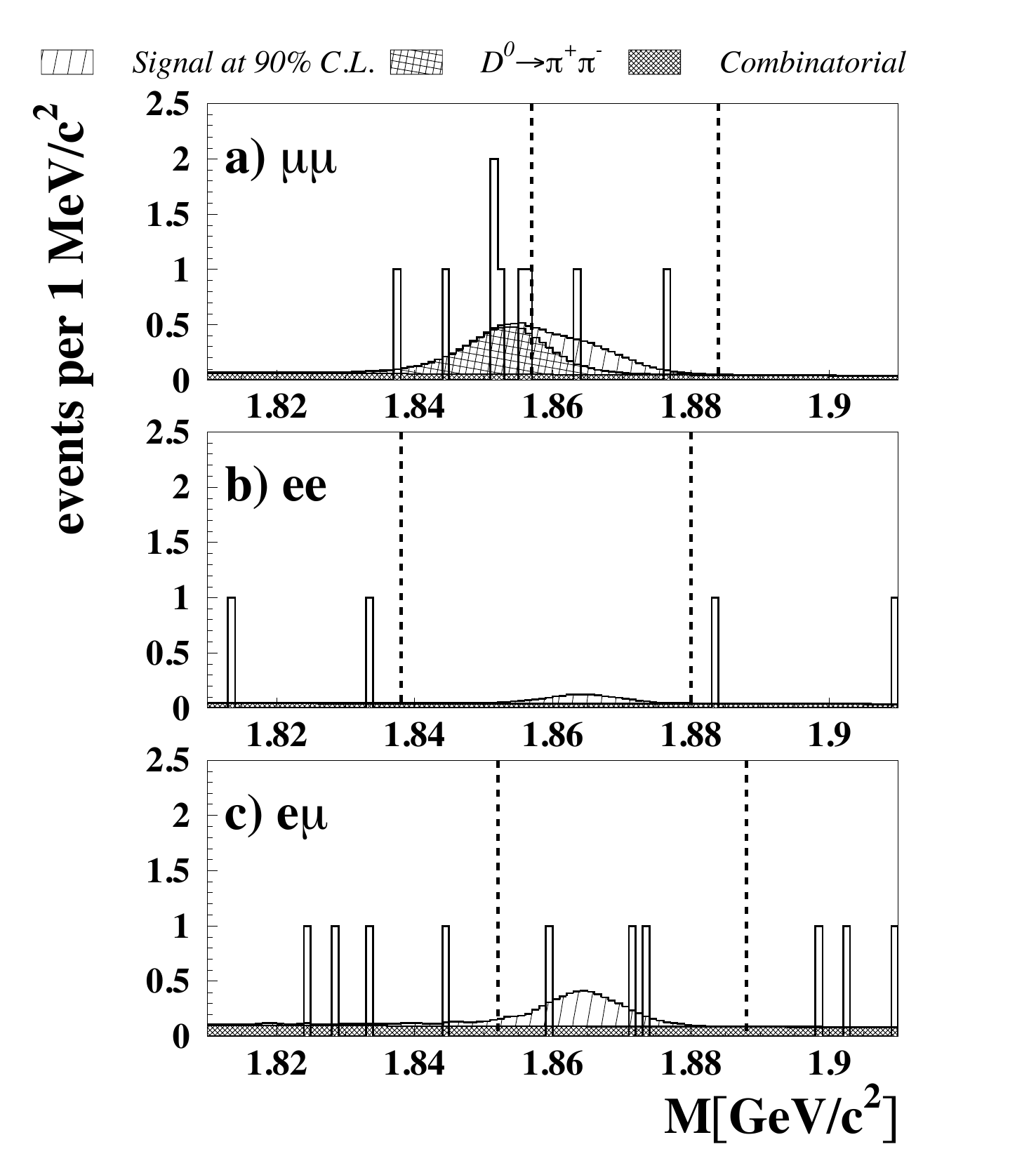}
\caption{The dilepton invariant mass distributions for (a) $\dz \to \mu^+\mu^-$, (b) $\dz \to e^+e^-$, and (c) $\dz \to e^{\pm}\mu^{\mp}$. The dashed vertical lines indicate the optimized signal window. Superimposed on the data (open histograms) are the estimated distribution for combinatorial background (filled histogram), the misidentification of $\dz~\to~\pi^+\pi^-$ (cross-hatched histogram), and the signal if the branching fractions were equal to the 90\% confidence level upper limit (single hatched histogram).
}
\label{fig:Dtoll} 
 \end{figure}

The invariant mass distributions after applying 
the optimized event selection criteria are shown in Fig.~\ref{fig:Dtoll}. In the signal region we find
two candidates in the $\dz \to \mu^+\mu^-$, zero candidates in the $\dz \to e^+e^-$, and
three candidates in the $\dz \to e^{\pm}\mu^{\mp}$ decay mode; the yields are consistent
with the estimated background.

\begin{figure}[htp]
\includegraphics[width=0.8\columnwidth]{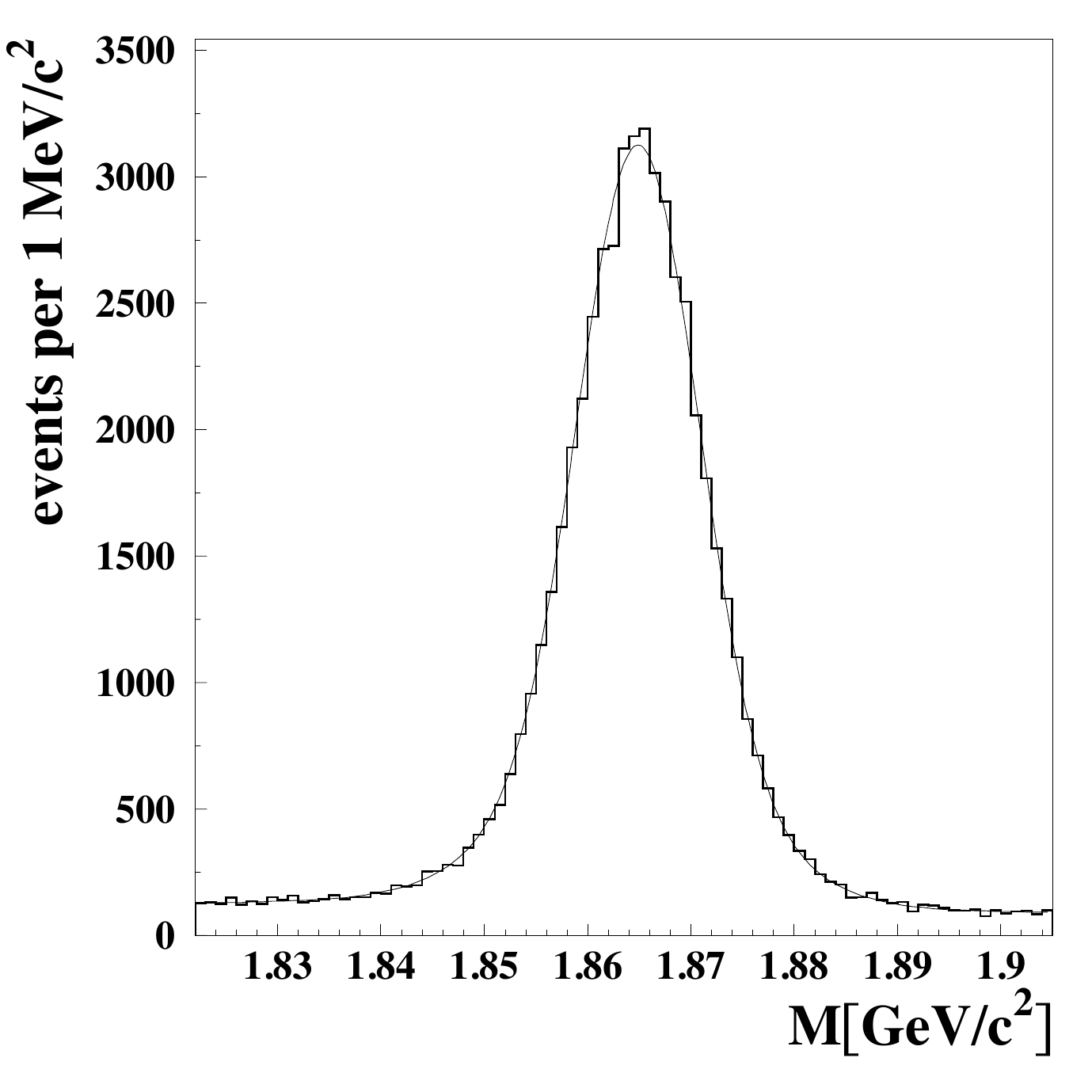}
\caption{The invariant mass distribution of $\dz \to \pi^{+}\pi^{-}$ with the fit superimposed using $\mu\mu$ selection criteria.
}
\label{fig:Dtopipi} 
 \end{figure}

A binned maximum likelihood fit is used to determine the yield of $\dz \to \pi^{+}\pi^{-}$ candidates for the normalization. We fit the invariant mass distribution using  the same kinematic selections as for individual leptonic
 modes, except for the criteria on $M$. The fit function is the sum
of two Gaussian distributions with the same mean and an FSR
 tail for the signal, and a first-order polynomial for the background. 
The shape of the FSR tail and its relative normalization are taken from the corresponding signal MC simulation.
The number of reconstructed $\dz$ mesons in the $\pi^{+}\pi^{-}$ mode is found to be $51.2\times10^{3}$, $44.1\times10^{3}$, and $46.0\times10^{3}$, using selection criteria for $\mu\mu$, $ee$, and $e\mu$ modes, respectively. The invariant mass distribution of $\dz \to \pi^{+}\pi^{-}$ using the $\mu\mu$ selection criteria with the fit curve superimposed is shown in Fig.~\ref{fig:Dtopipi}. The relative uncertainties on 
$N_{\pi\pi}$ are around 0.5\%.

The signal efficiencies are determined from the tuned signal MC simulation. In addition,
event weighting is applied to compensate for small differences in lepton and pion
identification efficiencies between data and MC simulation. The correction factors 
for lepton identification were obtained using $\gamma\gamma \to \ell^{+}\ell^{-}$ and $B \to X J/\psi (\to \ell^{+}\ell^{-})$ decays.
 The signal efficiencies are found to be between 5\% and 7\% for $\ell^+\ell^-$ decays and 
about 11\% for $\pi^+\pi^-$ decays. The uncertainties in $\epsilon_{\ell\ell}$
are estimated to be 0.3\% and include contributions from MC statistics (0.2\%),
lepton identification efficiency corrections (0.2\%), and MC tuning (0.1\%).
The uncertainty in $\epsilon_{\pi\pi}$ is smaller (0.1\%), because of a larger MC
sample, better known pion efficiency corrections and a negligible contribution from MC tuning, since a wider range in $M$ is used.

From the number of reconstructed $\dz \to \pi^+\pi^-$ decays, from the efficiency ratio, and from the known 
$\dz \to \pi^+\pi^-$ branching fraction the factors $f$
are calculated with Eq.~\ref{feq}. The relative uncertainties are around 5\% 
(see Table~\ref{tbl:br}) and include the errors on $N_{\pi\pi}$,
$\epsilon_{\ell\ell}$, $\epsilon_{\pi\pi}$ and the $\dz \to \pi^+\pi^-$ branching fraction, summed in quadrature.

Finally, the branching fraction upper limits (UL) are calculated using the program 
\texttt{pole.f}~\cite{conrad}, which extends
the Feldman-Cousins method~\cite{fc} by the inclusion of systematic uncertainties.
We find that the inclusion of systematic uncertainties produces nearly the same result as the standard Feldman-Cousins method.
The results are summarized in Table~\ref{tbl:br}. Note that $\mathcal{B}\left(\dz\to e^{\pm}\mu^{\mp}\right)$ denotes the sum $\brll{e}{\mu}+\brll{\mu}{e}$.

\begin{table}
  \caption{\label{tbl:br} Summary of the number of expected background events 
    ($N_{bkg}$), number of observed events ($N$) in the signal region, the reconstruction efficiencies ($\epsilon_{\ell\ell}$ and $\epsilon_{\pi\pi}$) of the $\dz\to\ell^{+}\ell^{-}$ and $\dz\to \pi^{+}\pi^{-}$ decays, the factors $f$ and 
    the branching fraction upper limits at the 90\% 
    confidence level.}
  \begin{tabular}{lccc}
    \hline
    \hline
    & $\dz\to \mu^+\mu^-$ & $\dz\to e^+e^-$ & $\dz\to e^\pm\mu^\mp$ \\
    \hline 
    $N_{bkg}$ & $3.1\pm 0.1$ & $1.7\pm0.2$ & $2.6\pm0.2$ \\
    $N$ & $2$ & $0$ & $3$ \\
    $\epsilon_{\ell\ell}[\%]$&$7.02\pm0.34$&$5.27\pm0.32$&$6.24\pm0.27$\\
    $\epsilon_{\pi\pi}[\%]$&$12.42\pm0.10$&$10.74\pm0.09$&$11.22\pm0.09$\\
    $f[10^{-8}]$ & $4.84(1\pm5.3\%)$ & $6.47(1\pm6.4\%)$ & $5.48(1\pm4.8\%)$ \\ 
    \hline
    UL  $[10^{-7}]$ & 1.4 & 0.79 & 2.6 \\
    \hline 
    \hline
  \end{tabular}
\end{table}

In summary, we have searched for the FCNC decays $\dz\to \mu^+\mu^-$ and
$\dz\to e^+e^-$, and the LFV decays $\dz\to e^\pm\mu^\mp$ 
using the Belle detector and have found no evidence of these decays. 
The upper limits on the branching fractions at the 90\% confidence level are
\begin{eqnarray}
\mathcal{B}\left(\dz\to \mu^+\mu^-\right)&<&1.4\times 10^{-7}\quad, \nonumber \\
\mathcal{B}\left(\dz\to e^+e^-\right)&<&7.9\times 10^{-8}\quad, \nonumber\\
\mathcal{B}\left(\dz\to e^\pm\mu^\mp \right)&<&2.6\times 10^{-7}\quad. \nonumber
\end{eqnarray}
Previously, the best upper limits on these decays were published by the BaBar 
Collaboration~\cite{babar} using 122~fb$^{-1}$ of data.
Our results improve these limits by a factor of 9 for $\dz\to \mu^+\mu^-$ decay, 
by a factor of 15 for $\dz\to e^+e^-$ decay, and by a factor of 3 for
$\dz\to e^\pm\mu^\mp$ decay. Recently, the CDF collaboration reported a preliminary
result on the UL for the $\dz\to \mu^+\mu^-$ branching fraction~\cite{CDF}; 
our result is lower by a factor of 3 and can further constrain the size of certain $R$-parity violating couplings. It also strongly disfavors a leptoquark contribution~\cite{Dorsner:2009cu} as the explanation for the anomaly in the measured  $\dsp\to \mu^+ \nu$ width~\cite{Fajfer:2009qg}.

\begin{acknowledgments}
We thank the KEKB group for the excellent operation of the
accelerator, the KEK cryogenics group for the efficient
operation of the solenoid, and the KEK computer group and
the National Institute of Informatics for valuable computing
and SINET3 network support.  We acknowledge support from
the Ministry of Education, Culture, Sports, Science, and
Technology (MEXT) of Japan, the Japan Society for the 
Promotion of Science (JSPS), and the Tau-Lepton Physics 
Research Center of Nagoya University; 
the Australian Research Council and the Australian 
Department of Industry, Innovation, Science and Research;
the National Natural Science Foundation of China under
Contracts No.~10575109, No.~10775142, No.~10875115, and No.~10825524; 
the Ministry of Education, Youth and Sports of the Czech 
Republic under Contracts No.~LA10033, and No.~MSM0021620859;
the Department of Science and Technology of India; 
the BK21 and WCU program of the Ministry of Education, Science and
Technology, National Research Foundation of Korea,
and NSDC of the Korea Institute of Science and Technology Information;
the Polish Ministry of Science and Higher Education;
the Ministry of Education and Science of the Russian
Federation and the Russian Federal Agency for Atomic Energy;
the Slovenian Research Agency;  the Swiss
National Science Foundation; the National Science Council
and the Ministry of Education of Taiwan; and the U.S.\
Department of Energy.
This work is supported by a Grant-in-Aid from MEXT for 
Science Research in a Priority Area (``New Development of 
Flavor Physics''), and from JSPS for Creative Scientific 
Research (``Evolution of Tau-lepton Physics'').
\end{acknowledgments}

\bibliography{Dtoll}

\begin{thebibliography}{20}
\bibitem [{cc()}]{cc}%
  \BibitemOpen
  \href@noop {} {}\bibinfo {note} {Throughout this paper charge-conjugate modes
  are included}\BibitemShut {NoStop}%
\end{thebibliography}

\end{document}